\documentclass[aps,pra,twocolumn,superscriptaddress]{revtex4-2}
\usepackage{amsmath,amsfonts,amsthm,bbm}
\usepackage[dvipsnames]{xcolor}
\usepackage{graphicx}
\usepackage{multirow,mhchem,chemformula}
\usepackage[colorlinks=true,breaklinks=true,linkcolor=red,citecolor=magenta,urlcolor=magenta]{hyperref}

\usepackage{tikz}
\usetikzlibrary{quantikz} 
\usepackage[english]{babel}
\usepackage[autostyle, english=american]{csquotes}
\usepackage{gensymb}
\MakeOuterQuote{"}
\usepackage{float}
\usepackage[markup=underlined]{changes}

\newcommand\crt[1]{\hat{a}^\dagger_{#1}}
\newcommand\dst[1]{\hat{a}^{\phantom{\dagger}}_{#1}}
\newcommand\bts[1]{{\bf{#1}}}

\usepackage{xcolor}

\usepackage{tikz} 
\newcommand*\circled[1]{\tikz[baseline=(char.base)]{
    \node[shape=circle,draw,inner sep=2pt] (char) {#1}}} 

\begin{document}

\title{Quantum-Centric Study of Methylene Singlet and Triplet States} 

\author{Ieva Liepuoniute*}
\affiliation{IBM Quantum, IBM Research - Almaden, 650 Harry Road, San Jose, CA 95120, USA}
\thanks{Corresponding author: ieva@ibm.com}
\author{Kirstin D. Doney}
\affiliation{Lockheed Martin, 3251 Hanover Street, Palo Alto, CA, 94304, USA}
\author{Javier Robledo Moreno}
\address{IBM Quantum, T. J. Watson Research Center, Yorktown Heights, NY 10598, USA}
\author{Joshua A. Job}
\affiliation{Lockheed Martin, 3251 Hanover Street, Palo Alto, CA, 94304, USA}
\author{Will S. Friend}
\affiliation{Lockheed Martin, 3251 Hanover Street, Palo Alto, CA, 94304, USA}
\author{Gavin O. Jones}
\affiliation{IBM Quantum, IBM Research - Almaden, 650 Harry Road, San Jose, CA 95120, USA}

\begin{abstract}
This study explores the electronic structure of the CH\(_2\) molecule, modeled as a (6e, 23o) system using a 52-qubit quantum experiment, which is relevant for interstellar and combustion chemistry. We focused on calculating the dissociation energies for CH\(_2\) in the ground state triplet and the first excited state singlet, applying the Sample-based Quantum Diagonalization (SQD) method within a quantum-centric supercomputing framework. We evaluated the ability of SQD to provide accurate results compared to Selected Configuration Interaction (SCI) calculations and experimental values for the singlet-triplet gap. To our knowledge, this is the first study of an open-shell system, such as the CH\(_2\) triplet, using SQD. To obtain accurate energy values, we implemented post-SQD orbital optimization and employed a warm-start approach using previously converged states.While the results for the singlet state dissociation were only a few milli-Hartrees from the SCI reference values, the triplet state exhibited greater variability. This discrepancy likely arises from differences in bit-string handling within the SQD method for open- versus closed-shell systems, as well as the inherently complex wavefunction character of the triplet state. The SQD-calculated singlet-triplet energy gap matched well with experimental and SCI values. This study enhances our understanding of the SQD method for open-shell systems and lays the groundwork for future applications in large-scale electronic structure studies using quantum algorithms.
\end{abstract}

\maketitle

\section{Introduction}

Accurate electronic structure calculations are essential for simulating molecules as they provide detailed descriptions of molecular energetics, reaction pathways, and spectroscopic properties~\cite{elliott2003chemistry, deimling1997radiation, jensen1971effects}. These quantum chemical calculations enable precise predictions of molecular behavior and spectroscopic properties, proving invaluable in understanding transient, radical, or toxic molecules that are difficult to measure experimentally~\cite{PuzzariniStanton2023, Zhao2014, Huang2011}). 
Classical quantum chemical calculation methods, such as coupled-cluster theory (CC) or density functional theory (DFT), while fundamental, can become computationally expensive for large, polyatomic molecules of chemical interest or may lack the necessary accuracy ~\cite{cohen2008insights, cohen2008fractional}. This is particularly true when addressing the high entanglement characteristic of open-shell radicals. Recent advances in quantum computing present new opportunities for effectively modeling these intricate systems~\cite{robledo2024chemistry, motta2024quantum, motta2024subspace, motta2023bridging}. Most notably, recent studies have demonstrated that electronic structure calculations of chemical relevance for multi-atom, and heavy atom molecules~\cite{robledo2024chemistry}, as well as supramulecular systems~\cite{kaliakin2024accurate} is now feasible using the Sample-based Quantum Diagonalization (SQD) technique within a Quantum-Centric Supercomputing (QCSC) framework~\cite{alexeev2024quantum, robledo2024chemistry, barison2024ext-SQD, kanno2023quantum, nakagawa2023adapt, kaliakin2024accurate}. The SQD technique has been successfully applied to electronic structure problems, including an iron-sulfur cluster (77 qubits) and hydrophobic interactions in a methane dimer (up to 54 qubits)~\cite{robledo2024chemistry,kaliakin2024accurate}, but its application to open-shell systems has not yet been explored. 

\begin{figure}[h]
\includegraphics[width=0.4\textwidth]{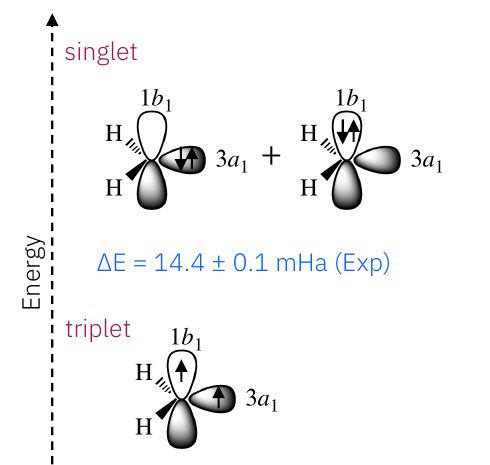}
\centering
\caption{Highest occupied molecular orbitals of methylene, along with the corresponding electronic configurations for the ground state triplet and the first excited state singlet. The experimentally determined singlet-triplet energy gap is ca. 14 m$\textrm{E}_h$ at $T_e$~\cite{jensen1988potential}.}
\label{figure:intro}
\end{figure}

Methylene (CH\(_2\)) is the prototype carbene and, as the smallest polyatomic free radical, is often used as a benchmark in different calculations to evaluate various theoretical methods~\cite{veis2010quantum, pittner1999assessment}. Additionally, CH\(_2\) is an important molecule in interstellar and combustion chemistry~\cite{schaefer1986methylene, herzberg1959spectrum, Bearda1992, jacob2021hunting}. From a computational perspective, CH\(_2\) serves as an important system for evaluating new methods, characterized by its triplet radical ground state (\(\tilde{X} \ ^3B_1\)) and the multireference nature of its lowest-lying excited state, a singlet non-radical (\(\tilde{a} \ ^1A_1\))~\cite{Harrison1974,Bunker1986,Gu2000,Chien2018, veis2010quantum}. The geometry and the equilibrium singlet-triplet energy gap (T\(_e\)) have been confirmed by extensive experimental and computational studies~\cite{Harrison1974,Bunker1986,Gu2000,Chien2018,hayden1982methylene,jordan1962lower,Petek1989,Kuchitsu2013, veis2010quantum}. For the two lowest states of CH\(_2\), the equilibrium geometry is \(C_{2v}\). The ground state is an open-shell triplet state, \(\tilde{X} \ ^3B_1\), which has a C-H bond length of ca. 1.09 \AA{} and a H-C-H angle of ca. 135.5\degree ~\cite{Kuchitsu2013}, with the electronic configuration \((1a_1)^2(2a_1)^2(1b_2)^2(3a_1)(1b_1)\). Conversely, the first excited state is a closed-shell singlet state, \(\tilde{a} \ ^1A_1\), which has a C-H bond length of 1.11 \AA{} and a H-C-H angle of ca. 102.4\degree ~\cite{Petek1989}, with two important electronic configurations: \((1a_1)^2(2a_1)^2(1b_2)^2(3a_1)^2\) and \((1a_1)^2(2a_1)^2(1b_2)^2(1b_1)^2\). The orbital occupation of the last two electrons defines the specific electronic state of methylene, as depicted in Figure \ref{figure:intro}. At the lowest point in the respective potential energy surfaces (PES), the energy gap between the two states T$_e$ is 14 m$\textrm{E}_h$ (equivalently 3159 cm$^{-1}$ or 9.03 kcal/mol)~\cite{jensen1988potential,Gu2000}.

Accurate classical calculations approaching the complete basis set (CBS) limit are needed to reproduce the experimental value of the singlet-triplet gap. Despite the efforts of various computational methods, significant discrepancies persist. The Hartree-Fock (HF) method, for example, grossly overestimates this gap due to the necessity of including a second configuration in the reference description of the $\tilde{a}$ $^1$A$_1$ state. This overestimation is also evident in the MP2 method, which relies on an insufficient single-configuration reference~\cite{schreiner1996carbene}. A two-configuration self-consistent field (SCF) treatment brings the gap down to approximately 17.5 m$\mathrm{E}_h$~\cite{yamaguchi1996x}. Furthermore, the gap shows a pronounced basis set effect, decreasing from 21.7 m$\textrm{E}_h$ with a double-zeta polarized (DZP) basis to 18 m$\textrm{E}_h$ with a triple-zeta polarized 2 (TZ2P) basis, and further to 17.4 m$\textrm{E}_h$ with a triple-zeta polarized 3 (TZ3P) basis. Expanding the basis set with additional diffuse functions and \(f\)-functions on carbon and \(d\)-functions on hydrogen reduces the gap by another 1 m$\mathrm{E}_h$. All \textit{ab initio} methods exhibit this basis set effect on the energy difference, with predictions ranging from 14 to 18 m$\textrm{E}_h$ when using at least a TZP basis set. Density functional theory (DFT) methods yield similar results, except for the B3PW91 functional \cite{das1999performance}. The most accurate calculation to date, utilizing the MRCISD/d-aug-cc-pV6Z method, predicts a gap that differs from experimental results by only 0.2 m$\mathrm{E}_h$~\cite{kalemos2004ch2}. 

In this study, we performed SQD calculations on the singlet and triplet states of CH\(_2\) using a correlation-consistent cc-pVDZ basis set, encompassing 6 electrons across 23 orbitals. This work represents the first open-shell analysis of molecular dissociation using SQD with a realistic correlation-consistent basis, and the first study of excited states on different $\textrm{U}(1)$ symmetry sectors. Additionally, this is the first study of a quantum phase transition that results from a level crossing using SQD. To assess the accuracy of the results, we compared them with Selected Configuration Interaction (SCI) calculations as implemented in \texttt{PySCF}~\cite{pyscf1, pyscf2}, and experimental energy values, providing insight into the capabilities of the SQD algorithm for both closed- and open-shell systems.

The structure of this work is as follows: first, we describe the methods employed, with a focus on large-scale noisy quantum hardware simulations and the associated error mitigation techniques. Next, we present and discuss the electronic energy results for the CH\(_2\) triplet and singlet states, and the calculation of the singlet-triplet gap using the SQD technique by simulating the C-H bond stretching. 

\section{Methods}

\begin{figure*}[t]
\centering
\includegraphics[width=0.85\textwidth]{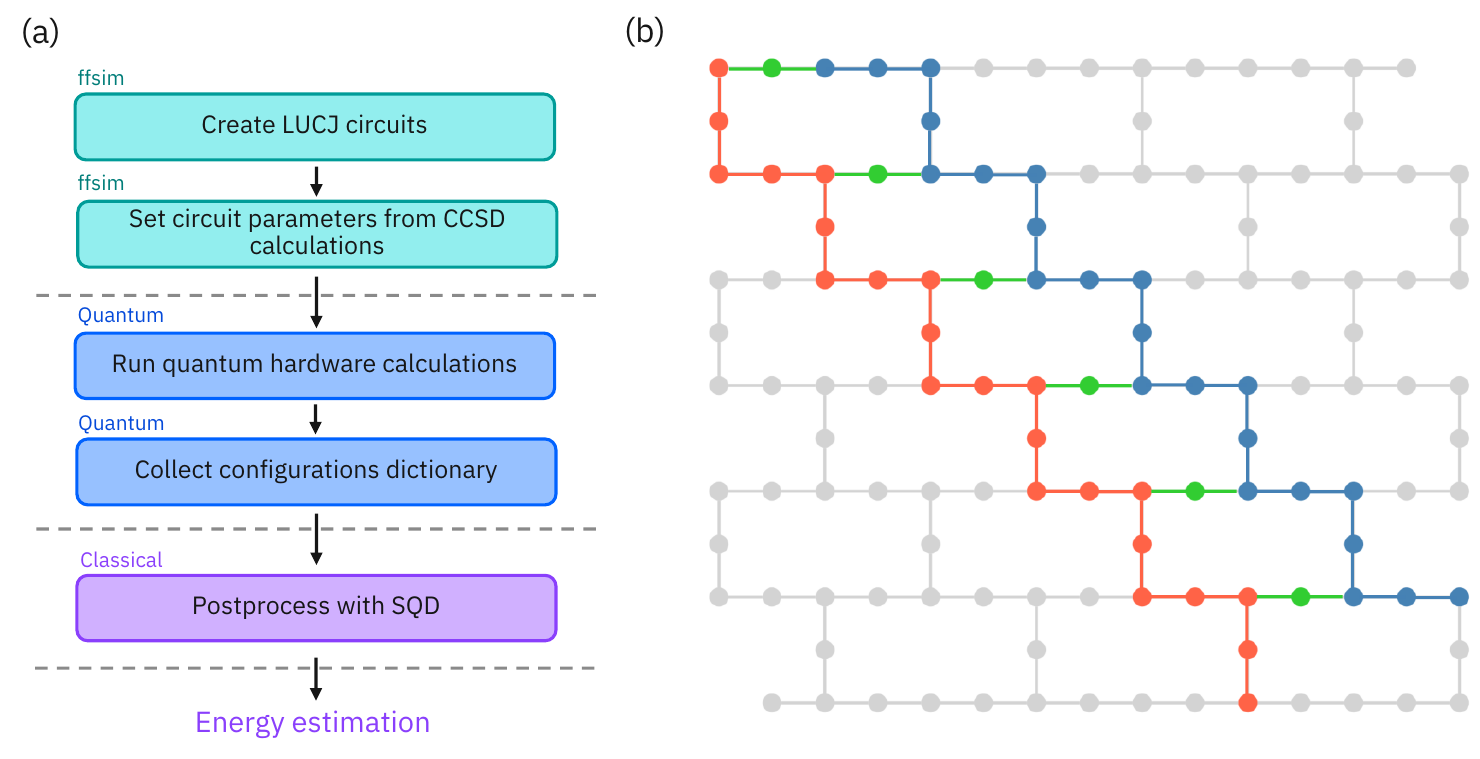}
\caption{Schematic representation of (a) the concerted approach workflow and (b) the qubit layouts of LUCJ circuits executed in this work for \((6e, 23o)\) methylene simulations using 52 qubits on \textit{ibm nazca}. Qubits used to encode the occupation numbers of \(\alpha\) (\(\beta\)) spin-orbitals are shown in red (blue). The auxiliary qubits mediate the
density-density interactions among orbitals of opposite spin are marked in green. Qubit layout was selected to maximize the number of ancilla qubits used to connect spin-up and spin-down orbitals.
}
\label{figure:methods}
\end{figure*}

\subsection{Classical Calculations}

\begin{figure}[h]
\includegraphics[width=0.45\textwidth]{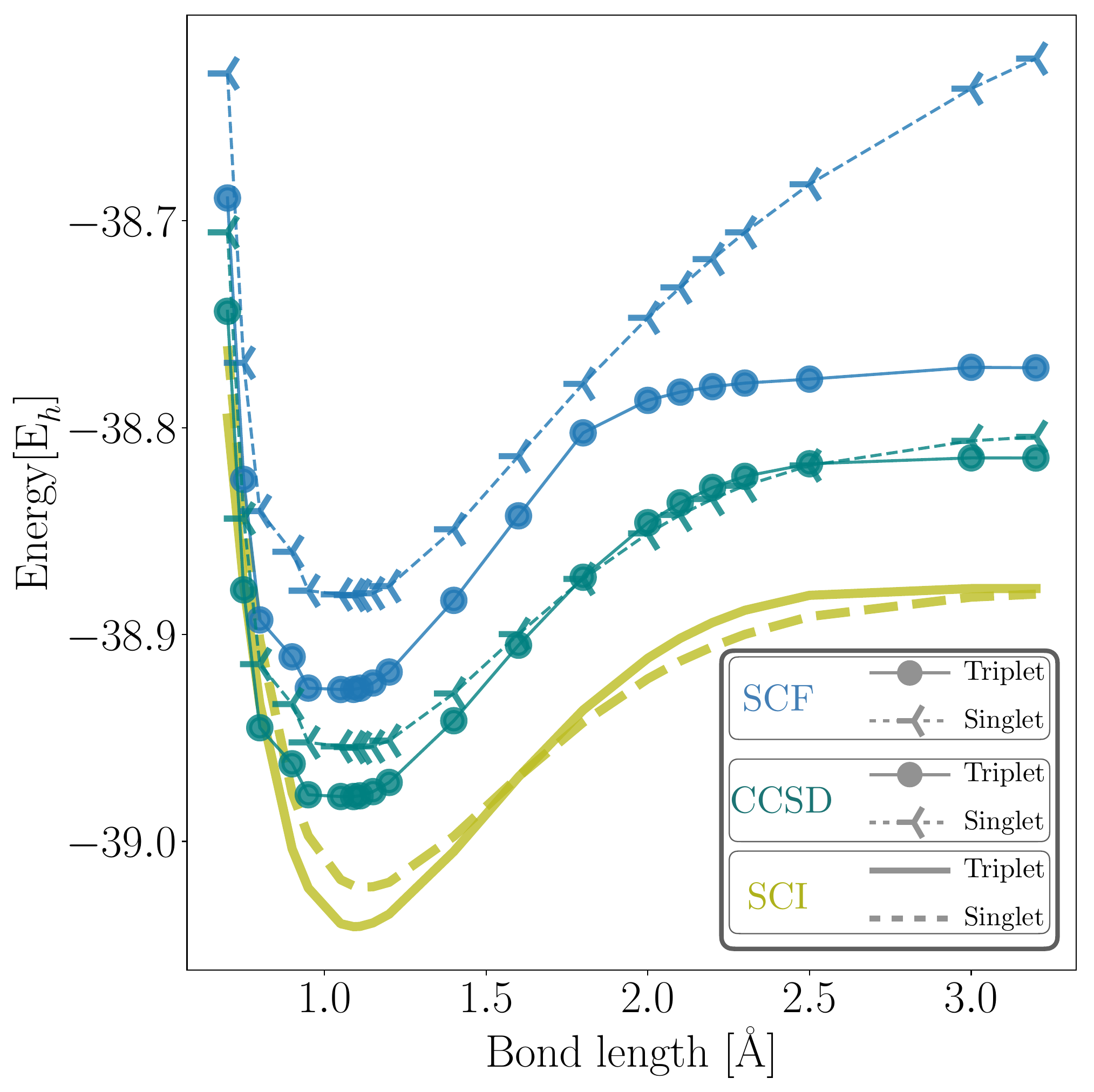}
\caption{Classical calculations for CH\(_2\) singlet and triplet dissociation profiles, performed using the cc-pVDZ basis set. The methods included self-consistent field (SCF), coupled cluster singles and doubles (CCSD), and Selected CI (SCI) with the frozen core approximation.
}
\label{figure:classical}
\end{figure}

Classical calculations were conducted using the open-source Python quantum chemistry package \texttt{PySCF}~\cite{pyscf1, pyscf2} to establish a reference for our quantum methods and to determine circuit parameters. The geometries for the CH\(_2\) singlet and triplet were obtained from experimentally determined data (as listed in the NIST CCCBDB database) and used in calculations without any further geometry optimization \cite{Petek1989,Kuchitsu2013,NIST}. As shown in Figure~\ref{figure:classical}, at equilibrium, the energy gap between the triplet and singlet states with the cc-pVDZ basis is approximately 18 m$\textrm{E}_h$ using SCI and 24 m$\textrm{E}_h$ using CCSD. These values are 3.5 and 9.5 m$\mathrm{E}_h$, respectively, from the experimental singlet-triplet gap. SCI, being the closest classical counterpart to SQD, is chosen as the method of reference to asses the accuracy of the quantum experiments.

The breaking of one of the C-H bonds presents a number of physically relevant features (Figure \ref{figure:classical}). First, at equilibrium, the triplet is the ground state, while the singlet is an excitation. Upon increasing the bond length, the roles of the singlet and triple are exchanged in a level crossing, leading to a phase transition of first order in the ground state wave function. In the large bond length limit, the singlet and triplet energies meet again. This last feature is easily explained by realizing that the electron in the isolated H atom is free to be oriented in any direction along its quantization axis. SCI accurately described these features, while CCSD fails to do so.

\subsection{Quantum Algorithms}

\subsubsection{Sample-based Quantum Diagonalization}

Sample-based Quantum Diagonalization ~\cite{kanno2023quantum,nakagawa2023adapt,robledo2024chemistry} is a 
variational approach for the search of eigenstates of many-body systems. The wave function ansatz is based on the expansion of a general many-body state in a subset $\mathcal{S}$ (of polynomial size in the number of electrons and orbitals) of the basis of single-particle electronic configurations:
\begin{equation}
    |\psi\rangle = \sum_{\bts{x} \in \mathcal{S}} \psi_{\bts{x}} |\bts{x}\rangle,
\end{equation}
where, $\bts{x} \in \{0, 1\}^{2N_\textrm{orb}}$ are bit-strings whose length is twice the number of spacial orbitals. The first and second halves of each bit-string represent the occupancy of the spin-up and spin-down orbitals respectively. $|\bts{x}\rangle$ represents a Slater determinant (or electronic configuration) of the form 
\begin{equation}
|\bts{x}\rangle = \prod_{p,\sigma} \left(\hat{a}_{p\sigma}^\dagger \right)^{x_{p\sigma}} |0\rangle, 
\end{equation}
where $p = 1, 2, ..., N_\textrm{orb}$, and $\sigma \in\{\uparrow, \downarrow \}$. The coefficients of the expansion $\psi_{\bts{x}}$ are defined by a lookup table. Provided that $|\mathcal{S}|$ grows polynomially with system size, their optimal values can be obtained by the diagonalization of the many-body Hamiltonian projected into the subspace $\mathcal{S}$:
\begin{equation}
\hat{H}_{\mathcal{S}}=\hat{\mathcal{P}}_{\mathcal{S}}\hat{H}\hat{\mathcal{P}}_{\mathcal{S}}
\;,
\end{equation}
where the projector $\hat{\mathcal{P}}_{\mathcal{S}}$ is
\begin{equation}
\hat{\mathcal{P}}_{\mathcal{S}} =\sum_{ {\bts{x}} \in \mathcal{S}} | {\bts{x}} \rangle \langle {\bts{x}} |.
\end{equation}

The size and constituents of $\mathcal{S}$ determine the accuracy of $|\psi\rangle$ to represent the target eigenstate. Classical heuristics exist to search the set of relevant configurations to include in $\mathcal{S}$, which go under the general umbrella of Selected Configuration Interaction methods~\cite{holmes2016heat, holmes2016efficient, smith2017cheap, sharma2017semistochastic}. Recent studies suggest that a quantum circuit $\Psi$ can also produce accurate statistical models $p(\bts{x})$ to produce samples that belong to $\mathcal{S}$~\cite{robledo2024chemistry} (SQD). In this second form, the quantum circuit $\Psi$ is prepared and measured in the computational basis. Under the Jordan-Wigner~\cite{jordan1993paulische} encoding, the sampled bit-strings are identified with the electronic configurations $\bts{x} \in \mathcal{S}$. Since the interacting electron problem preserves the total number of electrons and the number of spin-up and spin-down electrons separately, the Hamming weight of each half of any sampled bit-string must be equal to the number of spin-up and spin-down electrons. Therefore, circuits that are particle-number preserving are considered. See the following section for details about the specific form of the quantum circuit. 

Quantum noise modifies the distribution $p(\bts{x})$ into another $\tilde{p}(\bts{x})$, affecting the quality of the definition of the support of the variational ansatz. Consequently, the quantum device produces noisy samples that may break the particle-number symmetry. Ref.~\cite{robledo2024chemistry} introduces a self-consistent procedure to restore particle-number conservation using information from the diagonal of the one-body reduced density matrix $n_{p\sigma} = \langle \psi | \crt{p\sigma} \dst{p\sigma} | \psi \rangle$. In order to obtain better statistical convergence in the approximation of the ground state, at each configuration recovery iteration, $K$ batches of samples are considered, producing $K$ approximations to the ground state. The average orbital occupancy  for the configuration recovery step is the average value of $n_{p\sigma}$ over the $K$ batches of configurations. In this study, we take $K = 20$ for the triplet and $K = 10$ for the singlet.

Since the value of $n_{p\sigma}$ is not assumed to be known \textit{a priori} but computed and refined self-consistently, the first self-consistent recovery iteration consists on running the subspace projection and diagonalization by post-selecting the sampled bit-strings whose number of either spin-up or spin-down electrons is correct. Ten iterations of self consistent recovery are considered in this study.

The value of the total spin $S^2 = \langle \psi | \hat{S}^2|\psi \rangle$ is another conserved quantity in the many-electron problem. The conservation of this quantity can be improved by constructing $\mathcal{S}$ from the Cartesian product of all the spin-up and spin-down configurations on each batch, as detailed in Ref.~\cite{robledo2024chemistry}. Additionally, the addition of a soft constraint in the eigenstate solver improves the adherence to a fixed total spin sector of the Hilbert space:
\begin{equation}
    \left(\hat{H}_\mathcal{S} + \lambda \left[S^2 - s(s + 1) \right] \right) |\psi \rangle = E |\psi\rangle,
\end{equation}
where $\lambda$ is a Lagrange multiplier, whose value is set to $\lambda = 0.2$. 

In this work, we focus on the calculation of the ground and first excited state properties of the CH$_2$ molecule. In contrast to the recent approach proposed in Ref.~\cite{barison2024ext-SQD} to obtain excited states, the two eigenstates of interest live on different $\textrm{U}(1)$ symmetry sectors. The number of spin-up and spin-down electrons is the same in the singlet, while the triplet is obtained by flipping the spin of one of the electrons, resulting in a different number of spin-up and -down electrons. This observation can be exploited to cast the first excited state calculation as a ground state calculation in a different symmetry sector.

Orbital optimization is used as a post-SQD step to improve the accuracy of the results. The orbitals are optimized using gradient descent with the ADAM~\cite{kingma2017adammethodstochasticoptimization} optimizer, following the produce in Ref.~\cite{moreno2023enhancingexpressivityvariationalneural}. Some points in the dissociation curve where lower accuracy was initially reached, required warm starting the orbital optimization parameters from adjacent points in the dissociation curve that showed good accuracy.

The SQD method and orbital optimization were implemented using the \texttt{qiskit-addon-sqd}~\cite{qiskit-addon-sqd} python package.

\subsubsection{Local Unitary Cluster Jastrow (LUCJ) Ansatz}

We prepared our wavefunction guesses, used to approximate the ground state, from a truncated version of the Local Unitary Cluster Jastrow (LUCJ) ansatz~\cite{matsuzawa2020jastrow}. The corresponding LUCJ circuit was implemented with parameters derived from an efficient decomposition of the $T_2$ amplitudes computed from classical CCSD calculations. A potential decomposition of the 4-index cluster operator into a 2-index operator is achieved with cluster Jastrow ansatz via change in one particle basis. With an efficient parametrization for $T_2$ amplitudes of generalized qUCCSD, we obtain $\hat{e}^{\hat{T}- \hat{T}^\dagger} =\prod_l e^{\hat{K}^{(l)}} e^{\hat{J}^{(l)}} e^{-\hat{K}^{(l)}}$, where $l$ is the layer index, and $\hat{K}^{(l)}$ and $\hat{J}^{(l)}$ are one- and two-body operators, respectively:
\begin{equation}
    \begin{split}
        \hat{K}^{(l)} & = \sum_{pq,\sigma} K^{(l)}_{pq} \hat{a}^\dagger_{p\sigma}\hat{a}_{q\sigma} \\
        \hat{J}^{(l)} & = \sum_{pq, \sigma\tau} J^{(l)}_{pq, \sigma\tau} \hat{n}_{p\sigma} \hat{n}_{q\tau}.
    \end{split}
\end{equation}

\begin{figure*}[t]
\includegraphics[width=\textwidth]{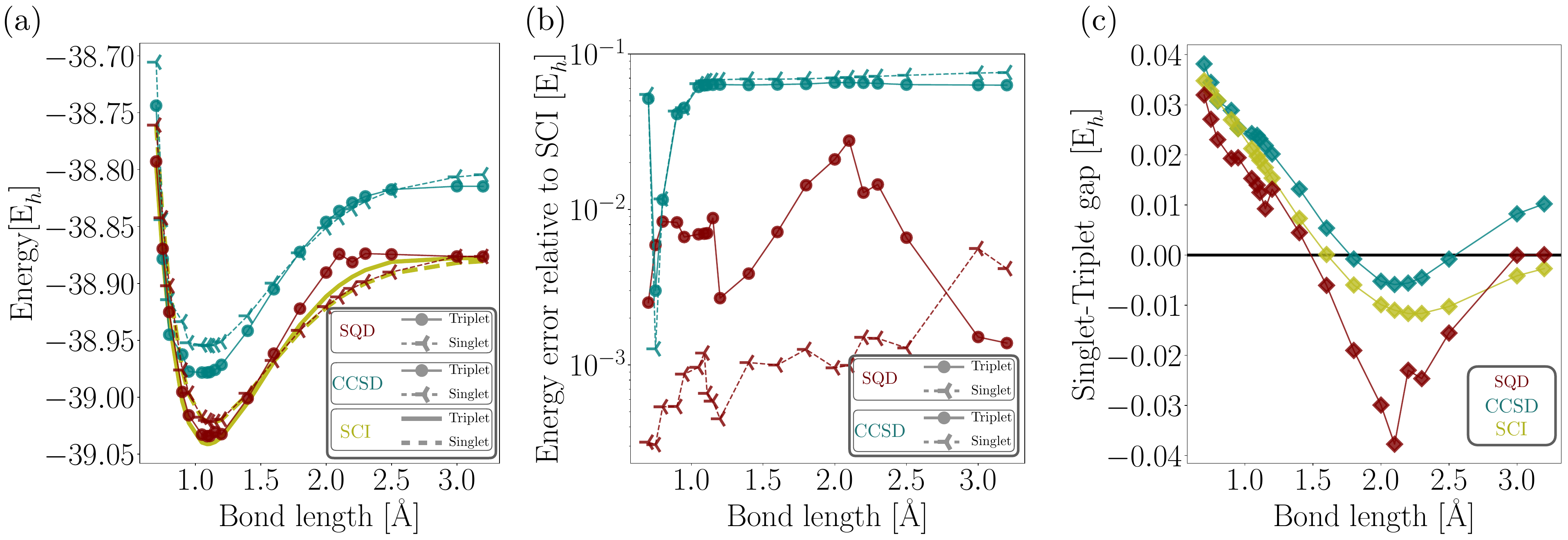}
\caption{(a) CH\(_2\) singlet and triplet dissociation energies along the potential energy surface (PES) for bond lengths ranging from 0.75 \, \text{Å} \text{ to } 3.20 \, \text{Å}, calculated using SQD, CCSD and SCI. (b) Energy error of SQD and CCSD relative to SCI. (c) Singlet-triplet gap as a function of C-H bond length for SQD, CCSD, and SCI calculations.}
\label{figure:triplet}
\end{figure*}

LUCJ is considered an efficient Variational Quantum Eigensolver (VQE) ansatz capable of reproducing the generalized Unitary Coupled Cluster ansatz for a sufficiently-large number of layers. Utilizing the Jordan-Wigner encoding, this ansatz can be realized on a quantum circuit without Trotter approximation. Each $e^{\hat{K}^{(l)}}$ can be exactly implemented by a Bogolyubov circuit ~\cite{google2020hartree}, \textit{i.e.}, a brickwork circuit of Givens rotations. Furthermore, each $e^{\hat{J}^{(l)}}$ can also be implemented exactly as it involves the exponentiation of commuting Pauli strings comprising only identity and $Z$ operations. This ansatz has been demonstrated to yield promising accuracy with reduced circuit depths compared to qUCCSD. 

In their study, Motta et al.~\cite{motta2023bridging} incorporated additional localization assumptions inspired by the Hubbard model on the $J^{(l)}_{pq, \sigma\tau}$ ``connectivity'', zeroing components of the tensor whose qubits are not adjacent. This integration results in a reduction in quantum computational resources and the ability to customize the approach for various hardware topologies. The depth for each $e^{\hat{K}^{(l)}}$ component is $\mathcal{O}(N_q)$, where $N_q$ is the number of qubits. The depth for each $e^{\hat{J}^{(l)}}$ component, with the locality assumptions is $\mathcal{O}(1)$. 

In this study the circuit parameters are obtained from the $t_1$ and $t_2$ coefficients from a CCSD calculation, as described in Ref.~\cite{robledo2024chemistry}. Two layers of the LUCJ ansatz were used. Quantum circuits were constructed using the \texttt{ffsim}~\cite{ffsim} software library, which enabled efficient simulation of fermionic quantum circuits and formed the foundation for steps 1 and 2 of the workflow in this study (Figure \ref{figure:methods} (a)).

\subsection{Quantum Hardware Calculations}

Quantum hardware calculations were conducted on \textit{ibm nazca}. After transpiling the circuits with the LUCJ ansatz for the \textit{ibm nazca} backend, we achieved a circuit depth of 1663, with a total of 3324 2-qubit gates. We submitted 40 jobs with various configurations corresponding to different bond distances along the dissociation profile for both the triplet and singlet states. Jobs consisted of 100,000 measurements (shots) for each circuit. Dynamical decoupling~\cite{pokharel2018demonstration, bollinger2009optimized, viola1998dynamical} and gate twirling~\cite{wallman2016noise} error mitigation techniques were employed to reduce noise originating from quantum gates. The qubit layout is depicted in Figure \ref{figure:methods}b. When mapped using the Jordan-Wigner transformation, this system was represented by 46 qubits—23 for \(\alpha\) and 23 for \(\beta\) spin-orbitals. Six additional qubits served as auxiliary qubits to mediate the density-density interactions between the spin-up and spin-down orbitals.

The LUCJ circuits were sampled in the computational basis to obtain a configuration dictionary containing bit-strings \(\bts{x} \in \{0, 1\}^{2N_{\textrm{orb}}}\) that represent electronic configurations (Slater determinants) distributed according to \(\tilde{p}(x)\). As shown in Figure \ref{figure:methods} (a), the collected configuration dictionary is then used in the SQD workflow.

\section{Results and Discussion}

\subsection{Energetics and Singlet-Triplet gap}

We first assessed the accuracy of the SQD method in comparison to the SCI and CCSD methods, alongside relevant experimental values. As illustrated in Figure \ref{figure:triplet} (a) and (b), the SQD estimates for the triplet energy are in good agreement with those obtained with SCI, with energy differences along the dissociation curve ranging from 1 to 28 m$\mathrm{E}_h$ and an average difference of approximately 7 m$\textrm{E}_h$ in the equilibrium region. Discontinuities and a deterioration of accuracy were observed in far-dissociation region, particularly between 2.00 \AA{} and 2.50 \AA{} (Figure \ref{figure:triplet} (a) and (b)), which can be attributed to the strongly correlated character of the triplet in that region of the dissociation curve. The SQD results for the CH\(_2\) singlet, on the other hand, remained within 1--4 m$\textrm{E}_h$ of the SCI reference values (Figure \ref{figure:triplet} (a) and (b)), demonstrating a remarkable level of agreement across all bond lengths.

Figure~\ref{figure:triplet}~(c) compares the singlet-triplet energy gap along the C--H bond dissociation pathway, as calculated using the SCI, CCSD, and SQD methods. Near equilibrium geometry, the gap values from all three methods are in close agreement. At this point, SQD yields a singlet-triplet energy gap of 19 m$\mathrm{E}_h$, closely matching the SCI-calculated gap of 18 m$\mathrm{E}_h$ with the cc-pVDZ basis set (Table~\ref{table:CH2-energies}). Although minor discrepancies exist between SQD and SCI results, favorable error cancellation in SQD brings their values into alignment. The experimental gap is 14 m$\mathrm{E}_h$, and achieving a closer match with experimental values rather than SCI would require larger basis sets. In contrast, the CCSD and SCF methods yield singlet-triplet gaps of 24 m$\mathrm{E}_h$ and 40 m$\mathrm{E}_h$, respectively. These results highlight SQD’s capability to produce energy gaps that are in strong agreement with both experimental benchmarks and SCI predictions, demonstrating robust performance.

\begin{table}[h]
\centering
\small
\caption{Comparison of singlet-triplet gap values obtained from various classical and quantum computational methods with experimental data}
\label{table:CH2-energies}
\begin{tabular*}{0.48\textwidth}{@{\extracolsep{\fill}}ccccc}
  \hline
  SCI & SQD & CCSD & SCF & Exp \\
  \hline
  18 m$\textrm{E}_h$& 19 m$\textrm{E}_h$& 24 m$\textrm{E}_h$& 40 m$\textrm{E}_h$& 14 m$\textrm{E}_h$\\
  \hline
\end{tabular*}
\end{table}
As the bond dissociates, the singlet-triplet energy gap approaches zero (Figure \ref{figure:triplet} (c)), signaling the first order phase transition in the ground state. This phenomenon is captured by all three methods, albeit at different critical bond lengths. The critical bond length obtained with SQD is in better agreement with the SCI value as compared to CCSD. Importantly, SQD accurately captures the vanishing of the triplet-singlet gap in the large bond-length limit. This behavior can be attributed to the fact that, at extended C--H bond distances, the electron on the dissociated hydrogen atom is effectively unbound and free to align or anti-align with the quantization axis. 

\subsection{Wavefunction 
Amplitude Analysis}

\begin{figure*}[t]
\centering
\includegraphics[width=0.99\textwidth]{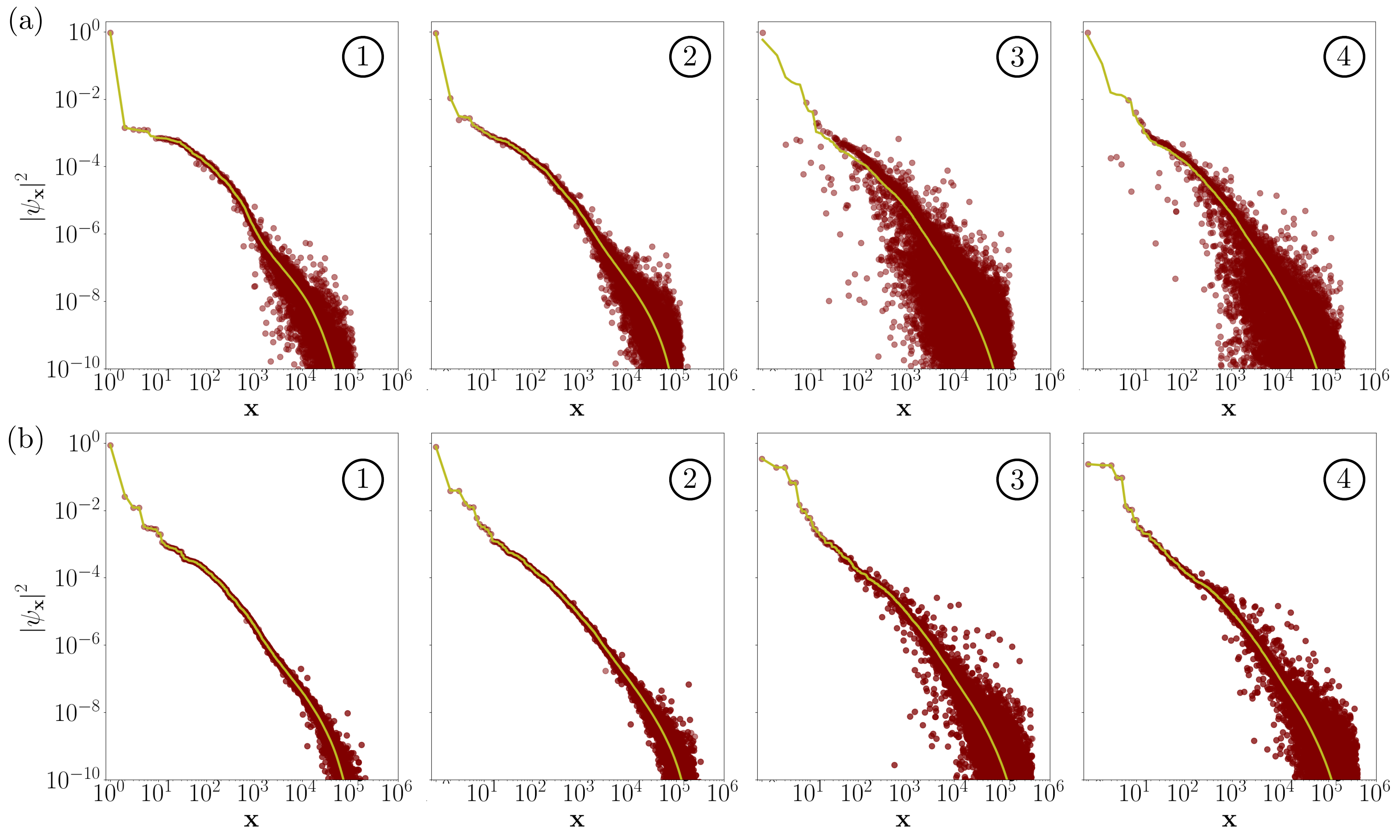}
\caption{Comparison of the SCI (yellow solid line) and SQD (brown dots) wavefunction amplitudes for all the electronic configurations $\mathbf{x} \in \mathcal{S}$. The configurations $\mathbf{x}$ are sorted along the horizontal axis in descending order of $\left|\psi_{\mathbf{x}}\right|^2$ of the SCI wavefunction. Panels (a) and (b) correspond to the singlet and triplet states, respectively, at bond lengths $1.11 \, \textrm{\AA} \rightarrow$ \protect\circled{1}; 
$1.60 \, \textrm{\AA} \rightarrow$ \protect\circled{2}; 
$2.30 \, \textrm{\AA} \rightarrow$ \protect\circled{3}; 
$2.50 \, \textrm{\AA} \rightarrow$ \protect\circled{4}.}
\label{figure:wavefunctions}
\end{figure*}

To better understand the challenges associated with triplet state calculations in the bond length range of 2.00 \AA{} to 2.50 \AA{}, we studied the accuracy of individual wavefunction amplitudes obtained from the SQD method, using the SCI method as a benchmark for comparison. Figure~\ref{figure:wavefunctions} shows a comparison between the SCI and SQD wavefunction amplitudes $\left| \psi_{\bts{x}}\right|^2$ for both the singlet (Figure \ref{figure:wavefunctions}a) and triplet (Figure \ref{figure:wavefunctions}b) states at various points along the dissociation curve. The agreement between SQD and SCI is stronger for larger wavefunction amplitudes and gradually deteriorates in the description of the tails of the wavefunction. Notably, at larger bond lengths, where higher levels of static correlation are more pronounced, the agreement in small wavefunction amplitudes deteriorates compared to those at bond lengths closer to the equilibrium geometry. This observation can be attributed to two key factors. First, at larger bond lengths, the wavefunction amplitudes are less concentrated compared to those near the equilibrium geometry. The second reason is that at bond lengths near the equilibrium geometry, the wavefunction exhibits a strong mean-field character, which diminishes at larger bond lengths. This directly impacts the shape of the reference spin-orbital occupancy that is used in configuration recovery. Specifically, when the wavefunction is dominated by mean-field effects, the reference occupancy resembles a sharp step function, with average occupancies of all spin-orbitals close to either 0 or 1. Static correlations can cause this step function to become smoother, resulting in some spin-orbitals having average occupancies that deviate from 0 or 1. The recovery of configurations in this study is more effective when the components of \( n \) are close to these values. 

In the SQD calculations for the CH\(_2\) triplet state, the multi-reference character is more pronounced for bond lengths between 2.00 \AA{} and 2.50 \AA{}, which contributes to the observed decrease in accuracy compared to the singlet state within the same range. Despite this, SQD produces accurate representations of the ground state wavefunction, even at the level of individual amplitudes. However, the challenges posed by less concentrated wavefunctions are notable; the accuracy of the SCI solver decreases, and configuration recovery becomes less effective due to the softer profiles of \(n\). Overall, this analysis underscores the performance of SQD while highlighting the specific difficulties encountered in this region.

\section{Conclusion}

In this study, we conducted electronic structure calculations for the two lowest-lying states of the CH\(_2\) molecule using quantum hardware. We specifically focused on the ground state triplet and the first excited state singlet, employing the SQD method within a quantum-centric simulation framework. This approach integrates quantum and classical calculations, enabling large-scale quantum computations and comprehensive post-processing of the data obtained from quantum hardware. Importantly, this research represents the first application of the SQD method to an open-shell molecule, particularly the CH\(_2\) triplet state.

We evaluated the effectiveness of the SQD algorithm for both open and closed-shell systems using a (6e, 23o) CH\(_2\) molecular system and conducting quantum hardware calculations on a 52-qubit system. The SQD results for the CH\(_2\) singlet state demonstrated excellent agreement with SCI results, deviating by only a few m$\mathrm{E}_h$. In contrast, the results for the triplet state were less consistent with SCI; however, they remained within a few m$\textrm{E}_h$ at equilibrium. The singlet-triplet energy gap showed close agreement with both SCI and experimental results, primarily due to beneficial error cancellation. However, in the triplet dissociation region, SQD struggled to capture the wavefunction character, an issue not observed with the singlet state. One potential extension to SQD could include new strategies for diversifying the ensemble of bit-strings in the open-shell case, such as forming compatible subsets of spin-up and spin-down subconfigurations, which may improve our ability to capture the character of the eigenvectors for open-shell systems.

Overall, this study enhances our understanding of the SQD method for both closed- and open-shell systems, laying the groundwork for future advancements and applications in accurate electronic structure studies using large-scale noisy quantum computers. In aerospace and defense applications, high-fidelity quantum chemical calculations play a critical role in modeling complex chemical environments. By advancing theoretical approaches, such as those benchmarked with the CH\(_2\) molecule, these simulations could enhance predictive models that can support the development of innovative sensing and detection technologies. The results in this work indicate that SQD has the potential to enable accurate electronic calculations for larger radical and transient species, as well as complex reactions relevant to combustion chemistry. Future work includes continued research on how SQD could expand the scale of what is possible with currently available and future hardware. As lower-noise and partially or fully error-corrected hardware becomes available, this technique, or extensions to it, may pave the way for early utility-scale applications and use cases in quantum computing.\\

\section{Acknowledgments}

We extend our sincere gratitude to everyone who supported this study. We especially thank our colleagues Mario Motta, Tanvi Gujarati, and Julia Rice for their valuable discussions, which greatly helped to refine our ideas and approach. We are also deeply grateful to Roberto Lo Nardo for his thorough reading of the manuscript and thoughtful feedback. Additionally, we are grateful to Caleb Johnson and Kevin J. Sung for their assistance with the SQD and ffsim software.

\bibliographystyle{iopart-num}
\bibliography{main}

\end{document}